\documentclass{article}
\usepackage{algorithm,algpseudocode}
\usepackage{amsfonts,amsmath,amssymb,amsthm,mathtools}
\usepackage{authblk}
\usepackage[scale=0.75]{geometry}
\usepackage{hyperref}
\usepackage{natbib}

\usepackage{enumitem}
\setlist[enumerate, 1]%
    {%
        leftmargin = \parindent,
        align = left,
        labelwidth=\parindent,
        parsep=0pt,
        labelsep = 2pt
    }

\newcommand{\calN}{\mathcal{N}}
\newcommand{\dd}{\mathrm{d}}

\newtheorem{remark}{Remark}

\title{Particle Gibbs without the Gibbs bit}
\author{Adrien Corenflos}
\affil[]{Department of Statistics, University of Warwick, UK}
\affil[]{\href{mailto:adrien.corenflos@warwick.ac.uk}{adrien.corenflos@warwick.ac.uk}}

\date{\today}

\begin{document}

\maketitle
\begin{abstract}
    Exact parameter and trajectory inference in state-space models is typically achieved by one of two methods: particle marginal Metropolis-Hastings (PMMH) or particle Gibbs (PGibbs). 
    PMMH is a pseudo-marginal algorithm which jointly proposes a new trajectory and parameter, and accepts or rejects both at once.
    PGibbs instead alternates between sampling from the trajectory, using an algorithm known as conditional sequential Monte Carlo (CSMC) and the parameter in a Hastings-within-Gibbs fashion.
    While particle independent Metropolis Hastings (PIMH), the parameter-free version of PMMH, is known to be statistically worse than CSMC, PGibbs can induce a slow mixing if the parameter and the state trajectory are very correlated.
    This has made PMMH the method of choice for many practitioners, despite theory  and experiments favouring CSMC over PIMH for the parameter-free problem.\
    Here, we describe a formulation of PGibbs which bypasses the Gibbs step, essentially marginalizing over the trajectory distribution in a fashion similar to PMMH.
    This is achieved by considering the implementation of a CSMC algortihm for the state-space model integrated over the \emph{joint distribution} of the current parameter and the parameter proposal.
    We illustrate the benefits of method on a simple example known to be challenging for PMMH.
\end{abstract}

\section{Introduction}\label{sec:intro}
This article is concerned with the problem of sampling from the joint distribution of models $\pi_T(x_{0:T}, \theta) \propto p(\theta) \gamma_T(x_{0:T} \mid \theta)$, defined recursively as
\begin{equation}\label{eq:non-Markovian-models}
    \begin{split}
        \pi_t(x_{0:t} \mid \theta) &= \frac{1}{Z_t(\theta)} \gamma_t(x_{0:t} \mid \theta) \\
        \gamma_t(x_{0:t} \mid \theta) &= p_t(x_t \mid x_{0:t-1}, \theta) g_t(x_{0:t}; \theta) \pi_{t-1}(x_{0:t-1} \mid \theta)
    \end{split}
\end{equation}
where, for $0 \leq t \leq T$, $g_t(x_{0:t}; \theta)$ is a potential function, $p_t(x_t \mid x_{0:t-1}, \theta)$ is a transition model, $p(\theta)$ is a prior distribution over the parameter $\theta$, and we have used the convention that $x_{s:t} = (x_s, \ldots, x_t)$, with $-1$ terms being ignored so that $\gamma_0(x_{0} \mid \theta) = p_0(x_0 \mid \theta) g_0(x_{0} \mid \theta) p(\theta)$.

A special case thereof, known as state-space models~\citep[see, e.g.,][]{cappe2005inference,sarkka2023bayesian}, arises when 
\begin{equation}\label{eq:ssm}
    \begin{split}
        g_t(x_{0:t}; \theta) &= p(y_t \mid x_t, \theta)
    \end{split}
\end{equation}
is the probability density function of a noisy observation $y_t$ conditionally on the state $x_t$ and the parameter $\theta$ and $p_t(x_t \mid x_{0:t-1}, \theta) = p(x_t \mid x_{t-1}, \theta)$ is a Markovian transition density.
State-space models are ubiquitous in the scientific and engineering literature, ranging from econometrics to ecology~\citep[see, e.g.,][for examples]{cappe2005inference,sarkka2023bayesian} and are often used to model time series data.

Another special case is given by Feynman--Kac models~\citep[see, e.g.,]{chopin2020book} 
\begin{equation}\label{eq:Feynman-Kac}
    \begin{split}
        \pi_t(x_{0:t} \mid \theta) &= \frac{1}{Z_t(\theta)} \gamma_t(x_{0:t} \mid \theta) \\
        \gamma_t(x_{0:t} \mid \theta) &= p_t(x_t \mid x_{t-1}, \theta) g_t(x_{t-1}, x_t; \theta) \pi_{t-1}(x_{0:t-1} \mid \theta)
    \end{split}
\end{equation}
which, too, have a Markovian transition model $p_t(x_t \mid x_{t-1}, \theta)$, but contrary to state-space models allow for bi-variate potential functions $g_t(x_{t-1:t}; \theta)$.
These appear naturally in scientific applications, where they are, for example, used to model the interaction between particles~\citep{delmoral2004book}, but also as computational constructions used to improve inference in state-space models~\citep{Pitt01061999}.

A key component to performing inference over the joint trajectory-parameter state space is the conditional $\pi_T(x_{0:T} \mid \theta)$ distribution, which is often intractable except in the simplest of cases~\citep[such as fully linear Gaussian state-space models][]{kalman1960}.
For state-space models, many density-assumed approximate methods exist, often leveraging approximate Gaussian conjugacy~\citep[see, e.g.,][for a comprehensive overview]{sarkka2023bayesian}.
However, these techniques are often limited in their applicability, as they require to postulate a tractable density family, and then assume that the conditional distribution $\pi_T(x_{0:T} \mid \theta)$ is ``close enough'' to being in that family: these assumptions are often not quantifiable, and bias arising from applying them is not easily controlled.

Feynman--Kac models, and, even more so, models of the form~\eqref{eq:non-Markovian-models} are doubly problematic in this regard, as they, in all generality, often do not even have approximate representations in known tractable families, and therefore, the use of assumed-density methods is not only difficult to control, but also often hardly justifiable.
Instead, inference within these is often conducted via simulation methods, taking the form of particle filters or similar sequential Monte Carlo (SMC) methods~\citep[see, e.g.,][]{delmoral2004book,chopin2020book}. 
At the heart of these methods lies the idea that the recursion~\eqref{eq:non-Markovian-models} can be propagated forward in time via successive proposal/comparison/selection steps, slowly building up a set of trajectories representing the posterior distribution $\pi_T(x_{0:T} \mid \theta)$.
The joint distribution $\pi_T(x_{0:T}, \theta)$ can then be sampled from in one of three ways: (i) jointly online, by means of a jittered particle filter over the trajectory and parameter space~\citep{chopin2013smc2,crisan2018nested}, (ii) jointly offline, by means of an algorithm called particle marginal Metropolis-Hastings (PMMH)~\citep{andrieu2010particle}, or (iii) alternatingly, sampling the trajectory and parameter in a Gibbs-like fashion, by means of an algorithm called particle Gibbs (PGibbs)~\citep{andrieu2010particle}.
In this article we focus on the latter two offline methods, and do not consider online methods further.

PMMH and PGibbs are both valid MCMC algorithms targeting the posterior distribution $p(x_{0:T}, \theta \mid y_{0:T})$, both using simulation techniques, but differ significantly in their \emph{modus operandi} and their properties.
On one hand, PMMH is a ``parameter-first'' algorithm, which tries to perform exploration over the marginal distribution $\pi_T(\theta) = \int \pi_T(x_{0:T}, \theta) \dd x_{0:T}$ using a Monte Carlo approximation of the normalizing constant and only recovers $\pi_T(x_{0:T} \mid \theta)$ as a byproduct.
In short, given a proposal $\theta' \sim q(\theta' \mid \theta)$, PMMH forms an unbiased Monte Carlo estimator $Z^N_T(\theta')$ of $Z_T(\theta')$ and then accepts or rejects the proposal with probability $\min(1, \frac{Z^N_T(\theta') q(\theta \mid \theta')}{Z^N_T(\theta) q(\theta' \mid \theta)})$.
On the other hand, PGibbs performs exploration by first sampling a new trajectory $x'_{0:T}$ from $\pi_T(x_{0:T} \mid \theta)$ via an MCMC kernel known as conditional Sequential Monte Carlo (CSMC), and then $\pi_T(\theta \mid x'_{0:T})$, in a Hastings-within-Gibbs-like fashion~\citep{geman1984stochastic,hastings1970mcmc}.
The two algorithms are therefore usually seen as complementary.
On the first hand PMMH is a pseudo-marginal algorithm~\citep{andrieu2009pseudomarginal} which targets the parameter posterior directly at the cost of some variance in the acceptance step.
PGibbs, on the other hand, leverages trajectory information directly, which is particularly interesting when the conditional $\pi_T(\theta \mid x_{0:T})$ is accessible in closed-form, or for which very performant MCMC methods exist.

In practice, however, the two algorithms are not equally performant: when the parameters and the trajectory are highly correlated, PGibbs can be slow to mix in view of well-established MCMC heuristics~\citep{papaspiliopoulos2007general}; conversly, for general models forming a marginal proposal distribution $q(\theta' \mid \theta)$ may be more complicated than leveraging information from the trajectory, in particular when the parameter prior is a conjugate prior to the state-space model (or close to be).

When it is fully conjugate, however, \citet{wigren2019parameter} shows that the parameter can be fully eliminated from the sampling procedure of PGibbs, so that marginal MCMC samples from $\pi_T(x_{0:T})$ are directly accessible, and therefore, $\pi_T(\theta) = \int \pi_T(\theta \mid x_{0:T}) \pi_T(x_{0:T}) \dd x_{0:T}$ can be marginally sampled too.
When only some parameters are conjugate, the marginalization can be combined with a PMMH intermediary step getting the both of both worlds. 

At the core of each method lies a distinct ``state-only'' algorithm: Particle independent Metropolis Hastings (PIMH) for PMMH, and CSMC for PGibbs.
A growing literature has been devoted to the theoretical properties of these two algorithms~\citep{chopin2015particle,andrieu2018uniform,lee2020coupled,karjalainen2023mixing}, and the consensus is that CSMC is a largely better algorithm than PIMH, in particular when the transition model $p_t(x_t \mid x_{0:t-1})$ can be evaluated and a ``backward sampling'' step~\citep{whiteley2010discussion} is used to regenerate the trajectory $x_{0:T}$ after a first selection stage.
Indeed, when backward sampling is used~\citep[at least for state-space models, see, e.g.,][]{lee2020coupled,karjalainen2023mixing} CSMC is stable under increasing number of observations, while PIMH is not, even when an additional backward sampling step is used therein~\citep[because this does not affect the acceptance function, see][Algorithm 3]{5940249}.

In view of this, a key question is whether PIMH can be replaced by CSMC as a marginal acceptance step for parameters.
In this article, we answer this in the positive: under the condition that the transition model (and the potential function) can be evaluated in closed form, Particle Gibbs can be marginalized, essentially skipping the ``Gibbs'' part of its namesake.
The article is organized as follows: (i) A general introduction to CSMC for non-Markovian models is given in Section~\ref{sec:csmc}, where we also introduce the notations used throughout the article, (ii) we then present our method in Section~\ref{sec:m-PGibbs}, where we show how to marginalize PGibbs, and (iii) we illustrate the method on a simple example in Section~\ref{sec:experiments}, where we show that the method dominates PMMH in terms of mixing properties.

\subsection{Notations and assumptions}\label{subsec:notations}

Given a population $x^{1:N} = \{x^1, \ldots, x^N\}$ of $N$ particles, we denote by $x^{-k} = \{x^1, \ldots, x^{k-1}, x^{k+1}, \ldots, x^N\}$ the collection of all particles except $x^k$.
The notation is extended to populations of trajectories: $x^{1:N}_{0:T} = \{x^1_{0:T}, \ldots, x^N_{0:T}\}$ is the collection of all trajectories.
When indexing over a specific trajectory, we will denote it as $x^{(n)}_{0:T}$, and this notation will be defined explicitly in context.
We also denote by $[N]$ the set of integers from $1$ to $N$ and $[N]_{-k} = \{1, \ldots, k, k+1, \ldots, N\}$ the set of integers £rom $1$ to $N$ except $k$.
When two subsets of the same trajectory $x_{s:t}$, $x_{t+1:u}$ are under consideration, we sometimes write $g(x_{s:t}, x_{t+1:u})$ to denote the function $g$ applied to the concatenation of the two sequences, i.e., $g(x_{s:t}, x_{t+1:u}) = g(x_{s:u})$.
This slight abuse of notation is used when indices for the subsets are not the same, and we want to avoid cluttering.
Throughout the article, we assume that the model $\pi_T(x_{0:T}, \theta)$ is such that the transition density $p_t(x_t \mid x_{0:t-1}, \theta)$ and the potential function $g_t(x_{0:t}; \theta)$ can be evaluated in closed form, and that the prior $p(\theta)$ is a proper which can be evaluated in closed form too.

Finally, while a short background on CSMC is provided in Section~\ref{sec:csmc} to ground notations, this article assumes some familiarity with the well-established version of the algorithm.
We refer the reader to the many tutorials and discussions on CSMC and PGibbs that have accumulated over the years, to name a few:
\citet[Chapter 5]{naesseth2019}; \citet{8187580}; \citet[Section 4]{martino_2018}; \citet[Chapter 3]{finke2015extended}, \citet[Chapter 16]{chopin2020book}; \citet[Chapter 4]{malory2021bayesian}; \citet{gauraha2020noteparticlegibbsmethod}; 
as well as the original article of~\citet{andrieu2010particle}.
\citet{naesseth2019,8187580} are particularly recommended as they take the direct non-Markovian route, which covers our method.

\section{Conditional sequential Monte Carlo for non-Markovian models}\label{sec:csmc}
Conditional sequential Monte Carlo~\citep{andrieu2010particle} is usually presented in the context of Markovian Feynman--Kac models.
Here, we present it for the more general context of non-Markovian systems~\eqref{eq:non-Markovian-models}.
This is because, even in the case of Markovian models, our marginalized method, presented in Section~\ref{sec:m-PGibbs} will formally be non-Markovian.

Consider now a non-Markovian model defined recursively as $\pi_t(x_{0:t}) = \frac{1}{Z_t} \gamma_t(x_{0:t})$ with 
\begin{equation}\label{eq:non-Markovian-models-recursion}
    \gamma_{t}(x_{0:t}) =  \eta_t(x_{0:t}) \pi_{t-1}(x_{0:t-1})
\end{equation}
for a decomposition $\eta_t(x_{0:t}) = p_t(x_t \mid x_{0:t-1}) g_t(x_{0:t})$ in terms of a transition density $p_t(x_t \mid x_{0:t-1})$ and a potential function $g_t(x_{0:t})$, similarly to the parametric model~\eqref{eq:non-Markovian-models}.
Given a number of particles $N > 1$, in its simplest form~\citep[using multinomial resampling]{gordon1993novel}, CSMC is a $\pi_T(x_{0:T})$-invariant Markov kernel $\mathcal{K}$ defined by Algorithm~\ref{alg:CSMC}.
\begin{algorithm}[t]
    \caption{CSMC kernel $\mathcal{K}(x'_{0:T} \mid x_{0:T})$}
    \label{alg:CSMC}
    \begin{algorithmic}[1]
        \State \textbf{Input:} $x_{0:T}$, $N$, backward sampling flag $B$.
        \State \textbf{Output:} $x'_{0:T}$.
        \State Set $x^{(1)}_{0:T} = x_{0:T}$, $a^{1}_{t} = 1$, for $t=0:T-1$.
        \State Sample $x^n_{0} \sim p_0(x_{0})$, for $n=2:N$ and set $x^{(n)}_{0} = [x^n_{0}]$.
        \For{$t=1:T$}
            \State Compute $W^n_{t-1} = \frac{g_{t-1}(x^{(n)}_{0:t-1})}{\sum_{m=1}^N g_{t-1}(x^{(m)}_{0:t-1})}$, for $n=1:N$.
            \For{$n=2:N$}
                \State Sample $a^n_{t-1} \sim \mathrm{Cat}(W^{1:N}_{t-1})$.
                \State Sample $x^n_{t} \sim p_t(x_{t} \mid x^{(a^n_{t-1})}_{0:t-1})$ and set $x^{(n)}_{0:t} = [x^{(a^n_{t-1})}_{0:t-1}, x^n_t]$.
            \EndFor
        \EndFor
        \State Compute $W^n_{T} = \frac{g_T(x^{(n)}_{0:T})}{\sum_{m=1}^N g_T(x^{(m)}_{0:T})}$, for $n=1:N$.
        \State Sample $k'_T \in \{2, \ldots, N\}$ with probability $\tfrac{W^{k'_T}_{T}}{1 - W^1_T}$
        \State Accept $k'_T$ with probability $\frac{W^1_T}{W^{k'_T}_{T}}$.
        \If{$B = 0$}
            \State Set $x'_{0:T} = x^{k'_T}_{0:T}$.
        \Else
            \State Sample $x'_T = x^{k'_T}_{T}$.
            \For{$t=T-1:0$}
                \State Compute $\tilde{w}^n_{t} = W^n_t \frac{\gamma_T(x^{(n)}_{0:t}, x'_{t+1:T})}{\gamma_{t}(x^{(n)}_{0:t})}$, for $n=1:N$.
                \State Sample $k_t \in \{1, \ldots, N\}$ with probability $\frac{\tilde{w}^{k_t}_{t}}{\sum_{m=1}^N \tilde{w}^{m}_{t}}$.
                \State Set $x'_{t:T} = [x^{k_t}_{t}, x'_{t+1:T}]$.
            \EndFor
        \EndIf
    \end{algorithmic}
\end{algorithm}
In this algorithm, $a^n_{t-1}$ is the ancestor of particle $n$ at time $t-1$, sampled as a categorical variable from the weights $W^{1:N}_{t-1}$.
The flag $B$ denotes whether a rejuvenation step is over the choice of ancestors for the final particle $x_T^{k_T}$.
When~\eqref{eq:non-Markovian-models} describes a state-space model, the backward sampling step has been shown to improve the mixing properties of the algorithm significantly~\citep{lee2020coupled,karjalainen2023mixing}.
In general, no simplification of $\frac{\eta_T(x^{(n)}_{0:t}, x'_{t+1:T})}{\eta_{t}(x^{(n)}_{0:t})}$ can be made, but, in the case of Feynman--Kac models~\eqref{eq:Feynman-Kac}, this ratio can be simplified to the incremental term $\gamma_t(x^n_{t-1}, x'_t)$, using the Markov property.

Finally, we note that many improvements over this algorithm can be made: for example, here, for simplicity of exposition, we used multinomial resampling, but other, low-variance resampling schemes can be used~\citep[see, e.g.,][]{chopin2015particle,karppinen2023bridge}.
Localization techniques can be used to improve the performance of the algorithm in high-dimensional settings~\citep{malory2021bayesian,finke2021csmc,corenflos2023auxiliary,corenflos2024particlemala} in a way resembling classical MCMC local exploration techniques.
Backward sampling can also be replaced by blocking strategies~\citep{singh2017blocking,karppinen2023bridge}.

Finally, when the model further depends on a parameter $\theta$, and given a proposal distribution $q(\theta' \mid \theta, x_{0:T})$, an additional step can be added to sample from $\pi_T(\theta \mid x_{0:T})$ as a Hastings-within-Gibbs procedure.
We do not describe this further, details being available in~\citet{andrieu2010particle}.

\section{Marginalizing Particle Gibbs}\label{sec:m-PGibbs}
\subsection{The target distribution}\label{subsec:target}
We now turn to the case of marginalizing PGibbs for Feynman--Kac models of the form~\eqref{eq:Feynman-Kac}.
To this end, let us assume that, rather than a proposal $q(\theta' \mid \theta, x)$ for the parameter $\theta$, we have a proposal $q(\theta' \mid \theta)$ which takes the form
\begin{equation}\label{eq:proposal-t-0}
    q(\theta' \mid \theta) = \int q_{\theta}(\theta' \mid u) q_u(u \mid \theta) \dd u
\end{equation}
for some auxiliary variable $u$ and a pair of (possibly different) proposal distributions $q_{\theta}(\theta' \mid u)$ and $q_u(u \mid \theta)$.
In the remainder of the article, we drop the subscripts for the sake of notational simplicity, and write $q(\theta' \mid u)$ and $q(u \mid \theta)$ instead.
This decomposition is key in the literature on methods for constructing proposals for ensemble MCMC algorithms~\citep{tjelmeland2004using}.
Examples of these include the case where $q(\theta' \mid u)$ and $q(u \mid \theta)$ are two halves of a Gaussian random walk proposal: $q(\theta' \mid u) = \mathcal{N}(\theta' \mid u, \tfrac{\delta}{2} I)$ and $q(u \mid \theta) = \mathcal{N}(u \mid \theta, \tfrac{\delta}{2} I)$, but also include pre-conditioned versions~\citep[see, e.g.,][]{titsias2018,corenflos2024particlemala} or could include more complex proposals combining a gradient oracle $g(\theta) \approx \nabla \log \pi(\theta)$ with a gaussian random walk to form a pseudo-Langevin proposal: $q(\theta' \mid u) = \mathcal{N}(\theta' \mid u, \tfrac{\delta}{2} I)$ and $q(u \mid \theta) = \mathcal{N}(u \mid \theta + \tfrac{\delta}{2} g(\theta), \tfrac{\delta}{2} I)$.

Consider now the joint distribution:
\begin{equation}\label{eq:augmented-joint-target}
    \pi(x_{0:T}, \theta^{1:M}, u, l) \propto \pi_T(x_{0:T}, \theta^l) q(u \mid \theta^l) \prod_{j \neq l} q(\theta^j \mid u).
\end{equation} 
This distribution describes the following object: (i) an index $l$ flags which of the parameters $\theta^{1:M}$ is at stationarity under $\pi_T(\theta)$, (ii) the trajectory $x_{0:T}$ is distributed according to $\pi_T(x_{0:T} \mid \theta^l)$, (iii) a variable $u$ ``decorrelates'' the $\theta^{1:M}$ and (iv) makes the $\theta^{-l}$ conditionally independent.

This distribution has the following conditionals:
\begin{equation}\label{eq:easy-conditionals-target}
    \begin{split}
        \pi(u \mid x_{0:T}, \theta^{l}) 
            &\propto q(u \mid \theta^l)\\
        \pi(\theta^{-l} \mid u, \theta^l, x_{0:T}) &= \prod_{j \neq l} q(\theta^j \mid u)
    \end{split}
\end{equation}
which are easy to sample from, while we can define $\pi(x_{0:T}, l \mid \theta^{1:M}, u)$ as the terminal distribution of the following sequence
\begin{equation}\label{eq:joint-feynman-kac-model}
    \begin{split}
        \pi_t(x_{0:t}, l \mid \theta^{1:M}, u) &= \frac{1}{Z_t(\theta^{1:M}, u)} \gamma_t(x_{0:t}, l \mid \theta^{1:M}, u) \\
        \gamma_t(x_{0:t}, l \mid \theta^{1:M}, u) &= p_t(x_t \mid x_{t-1}, \theta^l) g_t(x_{t-1}, x_{t}; \theta^l) \pi_{t-1}(x_{0:t-1}, l \mid \theta^{1:M}, u).
    \end{split}
\end{equation}

\subsection{Marginal CSMC target}\label{subsec:marg-csmc}
Under these notations, our goal is to form a CSMC algorithm for the marginalized distribution $\pi_T(x_{0:T} \mid \theta^{1:M}, u) = \sum_{l=1}^M \pi_T(x_{0:T}, l \mid \theta^{1:M}, u)$ corresponding to sampling from $x_{0:T}$, marginalizing over the choice of parameter $l$.
Once a marginal trajectory $x_{0:T}$ has been obtained, we can then sample the parameter $\theta^{1:M}$ from the distribution $\pi_T(\theta^{1:M} \mid x_{0:T})$ in closed form, or using a Hastings acceptance step.
To obtain the trajectory $x_{0:T}$ in the first place, we define the sequence of marginalized distributions
\begin{equation}\label{eq:marginal-feynman-kac-model}
    \begin{split}
        \pi_t(x_{0:t} \mid \theta^{1:M}, u) &= \frac{1}{Z_t(\theta^{1:M}, u)} \gamma_t(x_{0:t} \mid \theta^{1:M}, u) \\
        \gamma_t(x_{0:t} \mid \theta^{1:M}, u) &= \frac{\sum_{l=1}^M \gamma_t(x_{0:t}, l \mid \theta^{1:M}, u)}{\pi_{t-1}(x_{0:t-1} \mid \theta^{1:M}, u)}\pi_{t-1}(x_{0:t-1} \mid \theta^{1:M}, u).
    \end{split}
\end{equation}
This recovers the marginal distribution $\sum_{l=1}^M \pi_T(x_{0:T}, l \mid \theta^{1:M}, u)$ at the terminal time $T$.
Now, the ratio appearing in the definition of $\gamma_t(x_{0:t} \mid \theta^{1:M}, u)$ can be decomposed as
\begin{equation}
    \begin{split}
        \frac{\sum_{l=1}^M \gamma_t(x_{0:t}, l \mid \theta^{1:M}, u)}{\pi_{t-1}(x_{0:t-1} \mid \theta^{1:M}, u)} 
        &= \sum_{l=1}^M p_t(x_t \mid x_{t-1}, \theta^l) g_t(x_{t-1:t}; \theta^l) \pi_{t-1}(l \mid x_{0:t-1}, \theta^{1:M}, u)\\
        &\eqqcolon \eta_t(x_{0:t}; \theta^{1:M}, u).
    \end{split}
\end{equation}

\begin{remark}\label{rem:eta-non-Markovian}
    Even though the function $\eta_t$ depends on the full path $x_{0:t}$, it does so via the ``running'' posterior of $\theta$, which can be precomputed, so that the overall cost is commensurate to that of the classical particle Gibbs (times the number of $\theta$ particles $M$).
    Informally, the model is ``computationally'' Markovian.
\end{remark}
Now, given a choice of transition density $p_t(x_{t} \mid x_{0:t-1}, \theta^{1:M}, u)$, we can define the corresponding potential function
\begin{equation}\label{eq:potential-function}
    \begin{split}
        g_t(x_{0:t}; \theta^{1:M}, u, l') &\coloneqq \frac{\eta_t(x_{0:t}; \theta^{1:M}, u)}{p_t(x_{t} \mid x_{0:t-1}, \theta^{1:M}, u)}
    \end{split}
\end{equation}
which is, too, a function of the full path $x_{0:t}$ but, as per Remark~\ref{rem:eta-non-Markovian}, can be computed efficiently provided that the running posterior $\pi_{t-1}(l \mid x_{0:t-1}, \theta^{1:M}, u)$ is available and $p_t(x_{t} \mid x_{0:t-1}, \theta^{1:M}, u)$ can be evaluated easily.
We come back to the choice of $p_t(x_{t} \mid x_{0:t-1}, \theta^{1:M}, u)$ in Section~\ref{subsec:algo}.

In practice, once a new trajectory point $x_t$ has been sampled, the posterior $\pi_{t-1}(l \mid x_{0:t-1}, \theta^{1:M}, u)$ can be updated as
\begin{equation}\label{eq:posterior-update}
    \pi_{t}(l \mid x_{0:t-1}, \theta^{1:M}, u) = \frac{p_t(x_t \mid x_{t-1}, \theta^l) g_t(x_{t-1:t}; \theta^l) \pi_{t-1}(l \mid x_{0:t-1}, \theta^{1:M}, u)}{\sum_{m=1}^M p_t(x_t \mid x_{t-1}, \theta^m) g_t(x_{t-1:t}; \theta^m) \pi_{t-1}(m \mid x_{0:t-1}, \theta^{1:M}, u)}
\end{equation}
which is a simple re-weighting of the previous posterior $\pi_{t-1}(l \mid x_{0:t-1}, \theta^{1:M}, u)$.
This is a key point: both the state-space and the parameter space are updated in a single step, and the two are computationally feasible. 

\begin{remark}\label{rem:parameter-elimination}
    An interpretation of this section is as a special case of~\citet{wigren2019parameter}: 
    the index $l$ can be attributed a ``prior'' categorical distribution
    \begin{equation}\label{eq:l-prior}
        p(l \mid u, \theta^{1:M}) \propto p(\theta^l) q(u \mid \theta^l) \prod_{j \neq l} q(\theta^j \mid u)
    \end{equation}
    over $[N]$, which is updated at each time step as per~\eqref{eq:posterior-update}.
    Importantly (but trivially), the categorical distribution is conjugate to any likelihood model, so that PGibbs can be marginalized as per~\citet{wigren2019parameter}.
\end{remark}

\subsection{The algorithm}\label{subsec:algo}
Following Section~\ref{subsec:target} and Section~\ref{sec:csmc}, we now have a way, given $x_{0:T}, u, \theta^{1:M}, l$ to sample a new version of $\pi(x_{0:T} \mid u, \theta^{1:M})$ using a CSMC algorithm, after which we can sample a new parameter $\theta'$ from $\sum_{l=1}^M \pi_T(l \mid x_{0:T}, u, \theta^{1:M}) \delta(\theta^l)$.

Altogether, for a given choice of $q(u \mid \theta)$, $q(\theta' \mid u)$, and $p_t(x_t \mid x_{0:t-1}, \theta^{1:M}, u)$, this forms Algorithm~\ref{alg:m-PGibbs}, which we call \emph{marginalized PGibbs} (m-PGibbs).
\begin{algorithm}[t]
    \caption{Marginalized PGibbs (m-PGibbs) kernel}
    \label{alg:m-PGibbs}
    \begin{algorithmic}[1]
        \State \textbf{Input:} $x_{0:T}$, $\theta$, $l$, $N$, $M$.
        \State \textbf{Output:} $x'_{0:T}$, $\theta'$, $l'$.
        \State Sample $u \sim q(u \mid \theta)$.
        \State Sample $\theta^{-l} \sim \prod_{j \neq l} q(\theta^j \mid u)$.
        \State Sample $x'_{0:T}$ from Algorithm~\ref{alg:CSMC} targeting $\pi_T(x_{0:T} \mid u, \theta^{1:M})$ using a proposal $p_t(x_t \mid x_{0:t-1}, \theta^{1:M}, u)$ and the potential function $g_t(x_{0:t}; \theta^{1:M}, u, l')$ defined in~\eqref{eq:potential-function}.
        \State Sample $l' \in [N]$ with probability $\pi_T(l' \mid x_{0:T}, u, \theta^{1:M})$ where $\pi_T(l \mid x_{0:T}, u, \theta^{1:M})$ is defined as the terminal distribution of the sequence~\eqref{eq:joint-feynman-kac-model}.
    \end{algorithmic}
\end{algorithm}
In practice, in Algorithm~\ref{alg:m-PGibbs}, the quantity $\pi_T(l \mid x_{0:T}, u, \theta^{1:M})$ can be computed as a byproduct of the backward sampling step of the CSMC algorithm, by evaluating the incremental term appearing in~\eqref{eq:posterior-update} backwards in time.

The algorithm is a valid MCMC kernel targeting the joint distribution $\pi_T(x_{0:T}, \theta)$ as a collapsed Gibbs sampler~\citep{Liu01091994} for a marginalization of the auxiliary index $l$.
When $N\to \infty$, it is well known that the CSMC kernel becomes to an exact simulation of the target distribution $\pi_T(x_{0:T} \mid u, \theta^{1:M})$~\citep[Theorem 1]{andrieu2018uniform}.
The sampling of $x'_{0:T}$ can therefore be understood as being done marginally with respect to the distribution $\pi_T(l \mid u, \theta^{1:M})$, hence the name of the algorithm.
Under regularity conditions~\citep{chopin2015particle,andrieu2018uniform}, as $N \to \infty$, the algorithm recovers the following steps:
\begin{enumerate}
    \item Sample $u \sim q(u \mid \theta)$.
    \item Sample $\theta^{-l} \sim \prod_{j \neq l} q(\theta^j \mid u)$.
    \item Sample $x'_{0:T} \sim \pi_T(x_{0:T} \mid u, \theta^{1:M})$.
    \item Sample $l' \in [N]$ with probability $\pi_T(l' \mid x_{0:T}, u, \theta^{1:M})$.
\end{enumerate}
In particular, the last two steps taken together can be seen as sampling from $\pi_T(l' \mid u, \theta^{1:M})$, which, in turn, is a Barker acceptance step~\citep{barker1965monte} for the parameter $l$ if one were able to compute $\pi_T(l' \mid u, \theta^{1:M})$ in closed form.
This is to be compared with the interpretation of PMMH~\citep{andrieu2009pseudomarginal} as a noisy version of Metropolis--Hastings acceptance step for the same marginalized distribution.
Barker's acceptance step is known to be, at worst, half as efficient as Metropolis--Hastings acceptance steps~\citep{latuszynski2013clts}, but also more generally applicable~\citep{vats2021efficient}.
Sharper results on the relative inefficiency of Barker's acceptance step are available in the literature~\citep{Agrawal_Vats_Łatuszyński_Roberts_2023}, and suggest (in the high-dimensional limit) a more optimistic efficiency ratio of approximately $2/3$.

\subsection{Choice of the proposal mechanisms}\label{subsec:proposal}
Throughout this section, we have assumed that the parameter proposals $q(\theta' \mid u)$ and $q(u \mid \theta)$, as well as the transition density $p_t(x_t \mid x_{0:t-1}, \theta^{1:M}, u)$ where given.

Constructing the pair $(q(u \mid \theta), q(\theta' \mid u))$ is relatively straightforward, as discussed in Section~\ref{subsec:target}.
For example, a random walk proposal $q(\theta' \mid u) = \calN(\theta' \mid u, \tfrac{\delta}{2} \Sigma)$ and $q(u \mid \theta) = \calN(u; \theta, \tfrac{\delta}{2} \Sigma)$ is a natural choice for a pre-conditioning matrix $\Sigma$ and a step size $\delta$.
Calibrating both of these can then be achieved through classical adaptive methodology~\citep{haario2006dram,vihola2012robust}.
The choice of transition density $p_t(x_t \mid x_{0:t-1}, \theta^{1:M}, u)$ is more delicate and different configurations can result in different behaviours.

\begin{enumerate}
    \item \textbf{Prior mixture proposal:} A first choice is to set $p_t(x_t \mid x_{0:t-1}, \theta^{1:M}, u) = \sum_{m=1}^M p_t(x_t \mid x_{t-1}, \theta^{m}) \pi_{0}(m \mid \theta^{1:M}, u)$, that is to use the prior over $l$ to marginalize the transition density.
    \item \textbf{Posterior mixture proposal:} A second choice is to set $p_t(x_t \mid x_{0:t-1}, \theta^{1:M}, u, l) = \sum_{m=1}^M p_t(x_t \mid x_{t-1}, \theta^{m}) \pi_{t-1}(m \mid x_{0:t-1}, \theta^{1:M}, u)$, the mixture of the transition densities under the running posterior $\pi_{t-1}(m \mid x_{0:t-1}, \theta^{1:M}, u)$.
    \item \textbf{Closed-form prior mixture:} A third choice is to set $p_t(x_t \mid x_{0:t-1}, \theta^{1:M}, u) = \int p_t(x_t \mid x_{t-1}, \theta') q(\theta' \mid u) \dd \theta'$, that is to use the transition density $p_t(x_t \mid x_{t-1}, \theta^l)$ but marginalize over the proposal $q(\theta \mid u)$.
    This choice is only valid when the transition density $p_t(x_t \mid x_{t-1}, \theta)$ is conjugate to the proposal $q(\theta \mid u)$, and is not always possible.
\end{enumerate}
Both three choices are valid, but it is clear that the second choice is the most natural one, as the proposal will be adapted to the problem at hand rather than set at the start of the algorithm and we will consider this choice in the rest of the article.

Finally, when $q(u \mid \theta) = \calN(u \mid \theta, \tfrac{\delta}{2} \Sigma)$ and $q(\theta' \mid u) = \calN(\theta' \mid u, \tfrac{\delta}{2} \Sigma)$, it is worth noting that 
\begin{equation}\label{eq:proposal-identity}
    \begin{split}
        \pi_t(l \mid x_{0:t}, \theta^{1:M}, u) 
            &\propto \pi_t(x_{0:t} \mid \theta^l) p(l \mid u, \theta^{1:M}) \\
            &\propto \pi_t(x_{0:t} \mid \theta^l) p(\theta^l) q(u \mid \theta^l) \prod_{j \neq l} q(\theta^j \mid u)\\
            &\propto \pi_t(x_{0:t} \mid \theta^l) p(\theta^l) \prod_{j=1}^M q(\theta^j \mid u)\\
            &\propto \pi_t(x_{0:t} \mid \theta^l) p(\theta^l)
    \end{split}
\end{equation}
does not depend on $u$ but only on the parameters $\theta^{1:M}$.
This is a consequence of the fact that the proposal $q(u \mid \theta)$ is symmetric in $\theta$ and $u$, and that the prior $p(\theta)$ is independent of $u$ and is a common occurence in the literature using multi-try proposals~\citep[see, e.g., the discussion on RW-CSMC in][]{corenflos2024particlemala}.

\section{Empirical illustration}\label{sec:experiments}
In this section, we present the comparative performance of m-PGibbs against PMMH. 
An implementation is available at \url{https://github.com/AdrienCorenflos/m-PGibbs} to reproduce the results.
Because the focus is on their respective mixing properties, we do not compare to other algorithms such as PGibbs, or actively tune the parameters of the algorithms, including the Feynman--Kac model: all algorithms are run with a bootstrap particle filter. 
We use the simple case $M=2$ for the parameter $\theta$, which is in line with recent literature highlighting the limitations of ensemble methods~\citep{pozza2024fundamentallimitationsmultiproposalmarkov}.

We consider the same example as in~\citet[Section 16.5.1]{chopin2020book} as well as the same data.
This forms a simple state-space model with a Gaussian transition density $p_t(x_t \mid x_{t-1}, \rho, \sigma^2_X) = \calN(x_t \mid x_{t-1}, 0.1^2)$ and a Gaussian observation density $g_t(x_t) = \calN(y_t \mid x_t, \sigma^2_Y)$.
The parameter $\theta = (\rho, \sigma^2_X, \sigma^2_Y)$ is a vector of three parameters, and its prior is a product of three independent distributions: $p(\rho) = \mathcal{U}([-1, 1])$, $p(\sigma^2_X) = \mathcal{IG}(2, 2)$ and $p(\sigma^2_Y) = \mathcal{IG}(2, 2)$, an inverse gamma distribution.
The data $y_{0:T-1}$ is a vector of length $T=100$.

As explained in \citet[Section 16.5]{chopin2020book}, for a choice of $q(\theta' \mid \theta) = \mathcal{N}(\theta' \mid \theta, \tau I)$, with $\tau = 0.15^2$~\citep[this was calibrated in][to obtain approximately a 23.4\% acceptance rate of the parameter $\theta$ under the idealized MCMC chain]{chopin2020book}, PMMH only gets close to the optimal acceptance rate when $N \gg 1$.

To make PMMH and our method comparable, we take $q(u \mid \theta) = \mathcal{N}(u \mid \theta, \tfrac{\tau}{2} I)$ and $q(\theta' \mid u) = \mathcal{N}(\theta' \mid u, \tfrac{\tau}{2} I)$. 
This recovers $q(\theta' \mid \theta)$ marginally, and is a natural choice for the proposal.
The goal is to understand the behaviour of the algorithms as the number of particles $N$ increases.
Intuitively, we expect that CSMC will be able to explore the parameter space more efficiently than PMMH, for all values of $N$.
Because the proposals for the parameter $\theta$ are the same for all algorithms, it suffices to compare the acceptance rates of the individual samplers to compare them: indeed, a higher acceptance rate will result in a higher effective sample size~\citep{tierney1998note}.

\begin{figure}[t]
    \centering
    \includegraphics[width=0.7\textwidth]{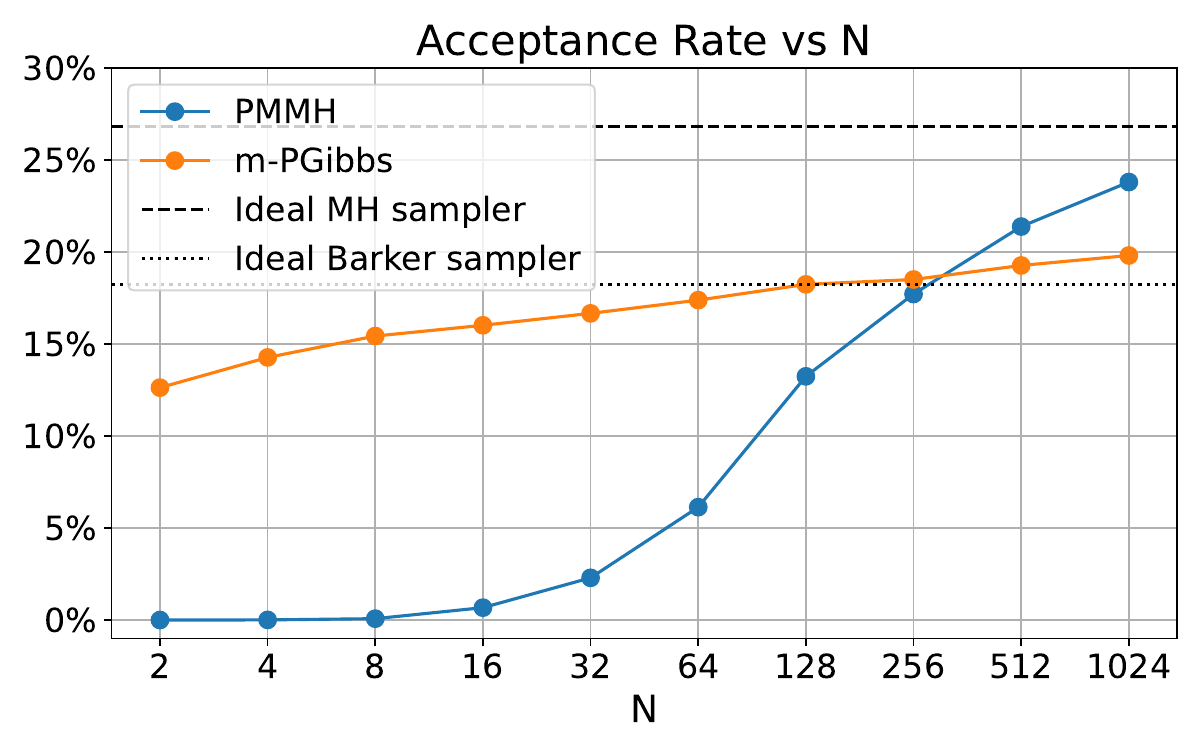}
    \caption{Comparison of m-PGibbs with posterior mixture proposal, and PMMH on the state-space model described in Section~\ref{sec:experiments}. 
    The figure shows the acceptance rate of the parameter $\theta$ as a function of the number of particles $N$.
    The dashed line indicates the realized ideal acceptance rate for PMMH, which is close to 27\%, simulated using a Kalman filter as a proposal in a classical Metropolis--Hastings step.
    The dotted line indicates the realized ideal acceptance rate for PMMH \emph{using a Barker's acceptance step}~\citep{barker1965monte}, which is 18\%, simulated using a Kalman filter too.}
    \label{fig:results}
\end{figure}

We ran all the algorithms for $100\,000$ iterations, and discarded the first $10\,000$ iterations as burn-in.
The results are shown in Figure~\ref{fig:results}.
Noting that PMMH is roughly twice as computationally efficient as all our m-PGibbs samplers for the same $N$, we see that m-PGibbs explores the parameter space more efficiently than PMMH for smaller values of $N$.
A point of note is that the acceptance rate of PMMH will always be lower than the ideal acceptance rate of Metropolis--Hastings~\citep{andrieu2009pseudomarginal} owing to its pseudomarginal nature, while m-PGibbs is not pseudomarginal and may improve on Barker's acceptance rate, as it does here.

For small values of $N$, PMMH simply fails to explore the parameter space, which is in line with prior analyzes of pseudomarginal methods~\citep{doucet2015efficient,sherlock2015efficiency}.
Eventually, for larger values of $N$, the acceptance rate of PMMH converges outperforms m-PGibbs: PMMH converges to the ideal acceptance Metropolis acceptance rate of approximately 27\%, while m-PGibbs converges to that of Barker's acceptance, which in this case is roughly 18\%: this is to be expected, as m-PGibbs implicitly uses Barker's acceptance step (see Section~\ref{subsec:algo}).
Still, m-PGibbs is largely more robust to the choice of $N$ than PMMH, which is a key point in favour of m-PGibbs.
Finally, we note that we also compared m-PGibbs with a ``classical'' PGibbs algorithm for the same proposal, sampling the parameter $\theta$ and the trajectory $x_{0:T}$ in two separate steps.
The results are not shown here, because PGibbs failed to mix properly, even for large values of $N$.

\section{Conclusion}\label{sec:conclusion}
This article has demonstrated that the PGibbs algorithm can be marginalized over the parameter space, forming a new algorithm called m-PGibbs.
Altogether, the proposed algorithm has a complexity cost of $O(T N M)$ in the number of time steps $T$, the number of state particles $N$ and the number of parameter particles $M$. 
While, for notational simplicity, we presented the algorithm for a general $M$, in practice we recommend taking $M=2$, resulting in a computational complexity of roughly twice that of PMMH.
This is due to the disappointing performance of ensemble MCMC samplers in the static regime~\citep{pozza2024fundamentallimitationsmultiproposalmarkov}.

We argue that \emph{if m-PGibbs can be implemented, it should usually be preferred to PMMH}. 
A caveat to this is the fact that m-PGibbs uses Barker's acceptance, which is less efficient than Metropolis--Hastings acceptance used in PMMH: provided one is willing to use a lot of particles, PMMH may eventually outperform m-PGibbs.
Augmenting the present method with a Metropolis--Hastings acceptance step (rather than Barker's) is likely possible via a careful consideration of the augmented distributions and proposals, but we have not explored this in detail.
While backed up by prior works on CSMC, this is a fairly strong statement, which we have so far only demonstrated up empirically in a simple example. 
However, due to the nature of the algorithm, a theoretical analysis of the algorithm in the style of~\citet{andrieu2018uniform,lee2020coupled,karjalainen2023mixing} is not straightforward and left for future work: for example, while PMMH can be proven to have a lower acceptance rate than its ``idealized'' version corresponding to $N \to \infty$~\citep{andrieu2009pseudomarginal}, there is no reason why this would be true for m-PGibbs (and it does not seem to be so empirically).
Still, we hope that this article further clarifies the picture of particle MCMC methods that has emerged over the past decade: as a building block, CSMC is a better algorithm than PIMH, and one should keep on building on it rather than defaulting to PIMH.

\section*{Acknowledgements}
The author thanks jetlag for the idea, and the resulting insomnia for the time to write this article.
Special thanks go Nicolas Chopin and Axel Finke for suggesting improvements to the manuscript, in particular to Axel for pointing out a mistake in the proposal mechanism which originally resulted in a biased Markov chain. 
The work was supported by UKRI grant EP/Y014650/1, as part of the ERC Synergy project OCEAN.

\bibliographystyle{apalike}
\bibliography{main}

\begin{thebibliography}{}

\bibitem[Agrawal et~al., 2023]{Agrawal_Vats_Łatuszyński_Roberts_2023}
Agrawal, S., Vats, D., Łatuszyński, K., and Roberts, G.~O. (2023).
\newblock {Optimal scaling of MCMC beyond Metropolis}.
\newblock {\em Advances in Applied Probability}, 55(2):492--509.

\bibitem[Andrieu et~al., 2010]{andrieu2010particle}
Andrieu, C., Doucet, A., and Holenstein, R. (2010).
\newblock Particle {M}arkov chain {M}onte {C}arlo methods.
\newblock {\em Journal of the Royal Statistical Society: Series B (Statistical Methodology)}, 72(3):269--342.
\newblock With discussion.

\bibitem[Andrieu et~al., 2018]{andrieu2018uniform}
Andrieu, C., Lee, A., and Vihola, M. (2018).
\newblock Uniform ergodicity of the iterated conditional {SMC} and geometric ergodicity of particle {Gibbs} samplers.
\newblock {\em Bernoulli}, 24(2):842--872.

\bibitem[Andrieu and Roberts, 2009]{andrieu2009pseudomarginal}
Andrieu, C. and Roberts, G.~O. (2009).
\newblock The pseudo-marginal approach for efficient {Monte Carlo} computations.
\newblock {\em The Annals of Statistics}, 37(2):697 -- 725.

\bibitem[Barker, 1965]{barker1965monte}
Barker, A.~A. (1965).
\newblock {M}onte {C}arlo calculations of the radial distribution functions for a proton--electron plasma.
\newblock {\em Australian Journal of Physics}, 18(2):119--134.

\bibitem[Cappé et~al., 2005]{cappe2005inference}
Cappé, O., Moulines, E., and Rydén, T. (2005).
\newblock {\em Inference in hidden {M}arkov models}.
\newblock Springer Series in Statistics. Springer.

\bibitem[Chopin et~al., 2013]{chopin2013smc2}
Chopin, N., Jacob, P.~E., and Papaspiliopoulos, O. (2013).
\newblock {SMC}\textsuperscript{2}: an efficient algorithm for sequential analysis of state space models.
\newblock {\em Journal of the Royal Statistical Society: Series B (Statistical Methodology)}, 75(3):397--426.

\bibitem[Chopin and Papaspiliopoulos, 2020]{chopin2020book}
Chopin, N. and Papaspiliopoulos, O. (2020).
\newblock {\em An Introduction to Sequential {M}onte {C}arlo}.
\newblock Springer.

\bibitem[Chopin and Singh, 2015]{chopin2015particle}
Chopin, N. and Singh, S.~S. (2015).
\newblock On particle {G}ibbs sampling.
\newblock {\em Bernoulli}, 21(3):1855--1883.

\bibitem[Corenflos and Finke, 2024]{corenflos2024particlemala}
Corenflos, A. and Finke, A. (2024).
\newblock {Particle-MALA and Particle-mGRAD: Gradient-based MCMC methods for high-dimensional state-space models}.
\newblock {\em arXiv preprint arXiv:2401.14868}.

\bibitem[Corenflos and S{\"a}rkk{\"a}, 2025]{corenflos2023auxiliary}
Corenflos, A. and S{\"a}rkk{\"a}, S. (2025).
\newblock {Auxiliary MCMC samplers for parallelisable inference in high-dimensional latent dynamical systems}.
\newblock {\em Electronic Journal of Statistics}, 19(1):1370 -- 1424.

\bibitem[Crisan and M{\'i}guez, 2018]{crisan2018nested}
Crisan, D. and M{\'i}guez, J. (2018).
\newblock {Nested particle filters for online parameter estimation in discrete-time state-space Markov models}.
\newblock {\em Bernoulli}, 24(4A):3039 -- 3086.

\bibitem[Del~Moral, 2004]{delmoral2004book}
Del~Moral, P. (2004).
\newblock {\em {F}eynman-{K}ac Formulae: Genealogical and Interacting Particle Systems with Applications}.
\newblock Springer.

\bibitem[Doucet et~al., 2015]{doucet2015efficient}
Doucet, A., Pitt, M.~K., Deligiannidis, G., and Kohn, R. (2015).
\newblock {Efficient implementation of Markov chain Monte Carlo when using an unbiased likelihood estimator}.
\newblock {\em Biometrika}, 102(2):295--313.

\bibitem[Finke, 2015]{finke2015extended}
Finke, A. (2015).
\newblock {\em On Extended State-Space Constructions for {M}onte {C}arlo Methods}.
\newblock PhD thesis, Department of Statistics, University of Warwick, UK.

\bibitem[Finke and Thiery, 2023]{finke2021csmc}
Finke, A. and Thiery, A.~H. (to appear, 2023).
\newblock Conditional sequential {Monte Carlo} in high dimensions.
\newblock {\em Annals of Statistics}.

\bibitem[Gauraha, 2020]{gauraha2020noteparticlegibbsmethod}
Gauraha, N. (2020).
\newblock {A Note on Particle Gibbs Method and its Extensions and Variants}.

\bibitem[Geman and Geman, 1984]{geman1984stochastic}
Geman, S. and Geman, D. (1984).
\newblock Stochastic relaxation, {Gibbs} distributions, and the {Bayesian} restoration of images.
\newblock {\em IEEE Transactions on pattern analysis and machine intelligence}, PAMI-6(6):721--741.

\bibitem[Gordon et~al., 1993]{gordon1993novel}
Gordon, N.~J., Salmond, D.~J., and Smith, A. F.~M. (1993).
\newblock Novel approach to nonlinear/non-{G}aussian {B}ayesian state estimation.
\newblock {\em {IEE} Proceedings F, Radar and Signal Processing}, 140(2):107--113.

\bibitem[Haario et~al., 2006]{haario2006dram}
Haario, H., Laine, M., Mira, A., and Saksman, E. (2006).
\newblock {DRAM}: efficient adaptive {MCMC}.
\newblock {\em Statistics and computing}, 16(4):339--354.

\bibitem[Hastings, 1970]{hastings1970mcmc}
Hastings, W.~K. (1970).
\newblock Monte {Carlo} sampling methods using {Markov} chains and their applications.
\newblock {\em Biometrika}, 57(1):97--109.

\bibitem[Kalman, 1960]{kalman1960}
Kalman, R.~E. (1960).
\newblock A new approach to linear filtering and prediction problems.
\newblock {\em Transactions of the ASME, Journal of Basic Engineering}, 82(1):35--45.

\bibitem[Karjalainen et~al., 2023]{karjalainen2023mixing}
Karjalainen, J., Lee, A., Singh, S.~S., and Vihola, M. (2023).
\newblock Mixing time of the conditional backward sampling particle filter.
\newblock {\em arXiv preprint arXiv:2312.17572}.

\bibitem[Karppinen et~al., 2023]{karppinen2023bridge}
Karppinen, S., Singh, S.~S., and Vihola, M. (2023).
\newblock Conditional particle filters with bridge backward sampling.
\newblock {\em Journal of Computational and Graphical Statistics}, 0(0):1--15.

\bibitem[{\L}atuszy{\'n}ski and Roberts, 2013]{latuszynski2013clts}
{\L}atuszy{\'n}ski, K. and Roberts, G.~O. (2013).
\newblock {CLTs} and asymptotic variance of time-sampled {M}arkov chains.
\newblock {\em Methodology and Computing in Applied Probability}, 15:237--247.

\bibitem[Lee et~al., 2020]{lee2020coupled}
Lee, A., Singh, S.~S., and Vihola, M. (2020).
\newblock Coupled conditional backward sampling particle filter.
\newblock {\em Annals of Statistics}, 48(5):3066--3089.

\bibitem[Lindsten and Schön, 2013]{8187580}
Lindsten, F. and Schön, T.~B. (2013).
\newblock {\em {Backward Simulation Methods for Monte Carlo Statistical Inference}}.

\bibitem[Liu, 1994]{Liu01091994}
Liu, J.~S. (1994).
\newblock {The Collapsed Gibbs Sampler in Bayesian Computations with Applications to a Gene Regulation Problem}.
\newblock {\em Journal of the American Statistical Association}, 89(427):958--966.

\bibitem[Malory, 2021]{malory2021bayesian}
Malory, S. (2021).
\newblock {\em Bayesian inference for stochastic processes}.
\newblock PhD thesis, Lancaster University.

\bibitem[Martino, 2018]{martino_2018}
Martino, L. (2018).
\newblock A review of multiple try {MCMC} algorithms for signal processing.
\newblock {\em Digital Signal Processing}, 75:134–152.

\bibitem[Naesseth et~al., 2019]{naesseth2019}
Naesseth, C.~A., Lindsten, F., and Sch\"on, T.~B. (2019).
\newblock Elements of sequential {Monte Carlo}.
\newblock {\em arXiv preprint arXiv:1903.04797}.

\bibitem[Olsson and Ryden, 2011]{5940249}
Olsson, J. and Ryden, T. (2011).
\newblock Rao-blackwellization of particle markov chain monte carlo methods using forward filtering backward sampling.
\newblock {\em IEEE Transactions on Signal Processing}, 59(10):4606--4619.

\bibitem[Papaspiliopoulos et~al., 2007]{papaspiliopoulos2007general}
Papaspiliopoulos, O., Roberts, G.~O., and Sk{\"o}ld, M. (2007).
\newblock {A General Framework for the Parametrization of Hierarchical Models}.
\newblock {\em Statistical Science}, 22(1):59 -- 73.

\bibitem[Pitt and and, 1999]{Pitt01061999}
Pitt, M.~K. and and, N.~S. (1999).
\newblock Filtering via simulation: Auxiliary particle filters.
\newblock {\em Journal of the American Statistical Association}, 94(446):590--599.

\bibitem[Pozza and Zanella, 2024]{pozza2024fundamentallimitationsmultiproposalmarkov}
Pozza, F. and Zanella, G. (2024).
\newblock {On the fundamental limitations of multiproposal Markov chain Monte Carlo algorithms}.

\bibitem[S{\"a}rkk{\"a} and Svensson, 2023]{sarkka2023bayesian}
S{\"a}rkk{\"a}, S. and Svensson, L. (2023).
\newblock {\em Bayesian filtering and smoothing}, volume~17.
\newblock Cambridge university press.

\bibitem[Sherlock et~al., 2015]{sherlock2015efficiency}
Sherlock, C., Thiery, A., and Golightly, A. (2015).
\newblock Efficiency of delayed-acceptance random walk {M}etropolis algorithms.
\newblock {\em arXiv preprint arXiv:1506.08155}.

\bibitem[Singh et~al., 2017]{singh2017blocking}
Singh, S.~S., Lindsten, F., and Moulines, E. (2017).
\newblock Blocking strategies and stability of particle {G}ibbs samplers.
\newblock {\em Biometrika}, 104(4):953--969.

\bibitem[Tierney, 1998]{tierney1998note}
Tierney, L. (1998).
\newblock A note on {M}etropolis--{H}astings kernels for general state spaces.
\newblock {\em Annals of applied probability}, pages 1--9.

\bibitem[Titsias and Papaspiliopoulos, 2018]{titsias2018}
Titsias, M.~K. and Papaspiliopoulos, O. (2018).
\newblock Auxiliary gradient-based sampling algorithms.
\newblock {\em Journal of the Royal Statistical Society: Series B (Statistical Methodology)}, 80(4):749--767.

\bibitem[Tjelmeland, 2004]{tjelmeland2004using}
Tjelmeland, H. (2004).
\newblock Using all {M}etropolis--{H}astings proposals to estimate mean values.
\newblock preprint 4/2004, Norwegian University of Science and Technology, Trondheim, Norway.

\bibitem[Vats et~al., 2021]{vats2021efficient}
Vats, D., Gonçalves, F.~B., Łatuszyński, K., and Roberts, G.~O. (2021).
\newblock {Efficient Bernoulli factory Markov chain Monte Carlo for intractable posteriors}.
\newblock {\em Biometrika}, 109(2):369--385.

\bibitem[Vihola, 2012]{vihola2012robust}
Vihola, M. (2012).
\newblock {Robust adaptive Metropolis algorithm with coerced acceptance rate}.
\newblock {\em Statistics and computing}, 22:997--1008.

\bibitem[Whiteley, 2010]{whiteley2010discussion}
Whiteley, N. (2010).
\newblock Discussion on particle {M}arkov chain {M}onte {C}arlo methods.
\newblock {\em Journal of the Royal Statistical Society: Series B}, 72(3):306--307.

\bibitem[Wigren et~al., 2019]{wigren2019parameter}
Wigren, A., Risuleo, R.~S., Murray, L., and Lindsten, F. (2019).
\newblock Parameter elimination in particle {Gibbs} sampling.
\newblock {\em Advances in Neural Information Processing Systems}, 32.

\end{thebibliography}

\end{document}